\documentclass[twocolumn]{revtex4}
\usepackage{graphicx}

\def\duzomniejsze{<\kern-.7mm<}
\def\duzowieksze{>\kern-.7mm>}

\def\textbf#1{{\bf #1}}
\def\beq{\begin{equation}}
\def\eeq{\end{equation}}
\def\be{\begin{equation}}
\def\ee{\end{equation}}
\def\ben{\begin{eqnarray}}
\def\een{\end{eqnarray}}
\def\beqa{\begin{eqnarray}}
\def\eeqa{\end{eqnarray}}
\def\eea{\end{array}}
\def\bea{\begin{array}}
\newcommand{\bei}{\begin{itemize}}
\newcommand{\eei}{\end{itemize}}
\newcommand{\bee}{\begin{enumerate}}
\newcommand{\eee}{\end{enumerate}}

\def\>{\rangle}
\def\<{\langle}

\begin{document}

\title{Generation of maximally entangled mixed states of two atoms via on-resonance asymmetric atom-cavity couplings}

\begin{abstract}
{\normalsize A scheme for generating the maximally entangled mixed
state of two atoms on-resonance asymmetrically coupled to a single
mode optical cavity field is presented. The part frontier of both
maximally entangled mixed states and maximal Bell violating mixed
states can be approximately reached by the evolving reduced
density matrix of two atoms if the ratio of coupling strengths of
two atoms is appropriately controlled. It is also shown that
exchange symmetry of global maximal concurrence is broken if and
only if coupling strength ratio lies between $\frac{\sqrt{3}}{3}$
and $\sqrt{3}$ for the case of one-particle excitation and
asymmetric coupling, while this partial symmetry-breaking can not
be verified by detecting maximal Bell violation.

}

\end{abstract}

\author{Shang-Bin Li}\email{stephenli74@yahoo.com.cn},

\affiliation{Shanghai research center of Amertron-global,
Zhangjiang High-Tech Park, \\
299 Lane, Bisheng Road, No. 3, Suite
202, Shanghai, 201204, P.R. China}

\maketitle

Quantum entanglement plays a crucial role in quantum information
processes \cite{Monroe}. In the last three years, much attention
has been paid to the characterization and preparation of the
maximally entangled mixed state
\cite{Peters2004,Barbieri2004,Ishizaka2000,Verstraete2001,Wei2003}.
Maximally entangled mixed states (MEMS) are those states that, for
a given mixedness, achieve the greatest possible entanglement. For
two-qubit systems and for various combinations of entanglement and
mixedness measures, the form of the corresponding MEMS has been
analyzed \cite{Wei2003}. At present, most of the experimental
verifications of the MEMS are based on polarized photons. The
cavity QED systems have been recognized as an important candidate
for implementing various kinds of quantum information processes,
and the precise control of the coupling of individual atoms to a
high-finesse optical cavity have been demonstrated experimentally
\cite{Guthohrlein,Eschner}. Current laboratory technologies have
demonstrated the feasibility of generating maximally entangled
state of two two-level atoms \cite{Zheng2000,Osnaghi2001}. For
generating the MEMS, the theoretical schemes based on the cavity
QED or collective interaction with environment have been proposed
\cite{Li2005,Clark2003,Li20051}. For entangling atoms,
on-resonance asymmetric coupling of individual atoms and cavity
field has exhibited certain advantage over the large-detuning
symmetric coupling because resonant cavity QED offers faster
entanglement schemes \cite{Olaya2004,Zhou2003}. This motivates us
to propose a feasible scheme for preparing the MEMS of two
two-level atoms on-resonance asymmetrically coupling to single
mode high-finesse optical cavity. It is found that the part
frontier of both MEMS and maximal Bell violating mixed states
(MBVMS) can be approximately reached by the evolving reduced
density matrix of two atoms if the ratio of coupling strengths of
two atoms is appropriately controlled. It is also shown that
exchange symmetry of global maximal concurrence is broken if and
only if coupling strength ratio lies between $\frac{\sqrt{3}}{3}$
and $\sqrt{3}$ for the case of one-particle excitation and
asymmetric coupling. However, the global maximum of maximal Bell
violation keep invariant under the exchange of two atoms, which
implies this partial symmetry-breaking can not be verified by
detecting maximal Bell violation. In the case of two-particle
excitation, the maximally entangled state of two atom can also be
generated when the coupling strength ratio is near 0.18. As the
ratio of coupling strengths tends to 1, the
critical-phenomenon-like behaviors of the global maximal
entanglement or Bell violation can be found.

The system discussed here consists of two two-level atoms confined
in a linear trap which has been surrounding by an optical cavity
\cite{Raimond2001}. We refer to atom 1 and atom 2. The Hamiltonian
describing the system is given by \cite{Plenio1999} ($\hbar=1$)
\beqa
H&&=\frac{\omega_1}{2}\sigma^{(1)}_z+\frac{\omega_2}{2}\sigma^{(2)}_z+\omega_aa^{\dagger}a\nonumber\\
&&+\lambda_1\theta_1(t)(a^{\dagger}\sigma^{(1)}_-+a\sigma^{(1)}_+)+\lambda_2\theta_2(t)(a^{\dagger}\sigma^{(2)}_-+a\sigma^{(2)}_+),
\eeqa where $a$ and $a^{\dagger}$ denote the annihilation and
creation operators for the single mode cavity field, and
$\sigma^{(i)}_z=|e\rangle_{ii}\langle{e}|-|g\rangle_{ii}\langle{g}|$,
$\sigma^{(i)}_+=|e\rangle_{ii}\langle{g}|$,
$\sigma^{(i)}_-=|g\rangle_{ii}\langle{e}|$ ($i=1,2$) are the
atomic operators. $\omega_1$ and $\omega_2$ are the atomic
transition frequencies of atom 1 and atom 2, respectively.
$\lambda_1$ and $\lambda_2$ are the coupling strengths between the
cavity field and atom 1, atom 2, respectively. $\theta_i(t)$
($i=1,2$) represent the time dependence of the atom-cavity
coupling. Plenio et al. have discussed the generation of maximally
entangled states in such a system \cite{Plenio1999}, in which the
cavity decay is continuously monitored. Here we investigate two
two-level atoms resonantly coupling with one mode optical cavity,
i.e. $\omega_1=\omega_2=\omega_a=\omega$ and
$\theta_1(t)=\theta_2(t)=\theta(t)$.

We assume that the initial state of the system (1) is described by
the density matrix
$\rho(0)=|0\rangle\langle0|\otimes|eg\rangle\langle{eg}|$, i.e.,
atom 1 is in the excited state, atom 2 is in the ground state and
the cavity field is in vacuum state, respectively. Substituting
$\rho(0)$ into the Schr\"{o}dinger equation, and tracing out the
degree of freedom of the cavity field, we could obtain the reduced
density matrix $\rho_a(t)$ describing the time evolution of two
atoms, \beqa
\rho_a(t)&=&\frac{\lambda^2_1}{2\lambda^2}(1-\cos2{\Theta(t)})|gg\rangle\langle{gg}|\nonumber\\
&&+\frac{\lambda^2_1}{2\lambda^2}(1+\cos2{\Theta(t)})|B_2\rangle\langle{B_2}|\nonumber\\
&&+\frac{\lambda^2_2}{\lambda^2}|B_1\rangle\langle{B_1}|\nonumber\\
&&-\frac{\lambda_1\lambda_2}{\lambda^2}\cos{\Theta(t)}(|B_1\rangle\langle{B_2}|+|B_2\rangle\langle{B_1}|),
\eeqa where,
$|B_1\rangle=\frac{1}{\lambda}(\lambda_1|ge\rangle-\lambda_2|eg\rangle)$,
$|B_2\rangle=\frac{1}{\lambda}(\lambda_1|eg\rangle+\lambda_2|ge\rangle)$,
and $\lambda=\sqrt{\lambda^2_1+\lambda^2_2}$ and
$\Theta(t)=\lambda\int^{t}_{0}\theta(\tau)d\tau$.

In order to quantify the degree of entanglement, we adopt the
concurrence to calculate the entanglement between two atoms. The
concurrence related to the density operator $\rho$ of a mixed
state is defined by \cite{Wootters1998} \be
C(\rho)=\max\{\delta_1-\delta_2-\delta_3-\delta_4,0\}, \ee where
the $\delta_i$ ($i=1,2,3,4$) are the square roots of the
eigenvalues in decreasing order of magnitude of the "spin-flipped"
density operator $R$ \be
R=\rho(\sigma_y\otimes\sigma_y)\rho^{\ast}(\sigma_y\otimes\sigma_y),
\ee where the asterisk indicates complex conjugation.

The explicit expression of the concurrence $C(t)$ characterizing
the entanglement of two atoms can be found to be \beqa
C(t)&=&|\frac{\lambda^3_1\lambda_2-2\lambda_1\lambda^3_2}{\lambda^4}+\frac{2\lambda_1\lambda_2(\lambda^2_2-\lambda^2_1)}{\lambda^4}\cos{\Theta(t)}\nonumber\\
&&+\frac{\lambda^3_1\lambda_2}{\lambda^4}\cos2{\Theta(t)}|, \eeqa
where $|x|$ gives the absolute value of $x$.

\begin{figure}
\centerline{\includegraphics[width=2.5in]{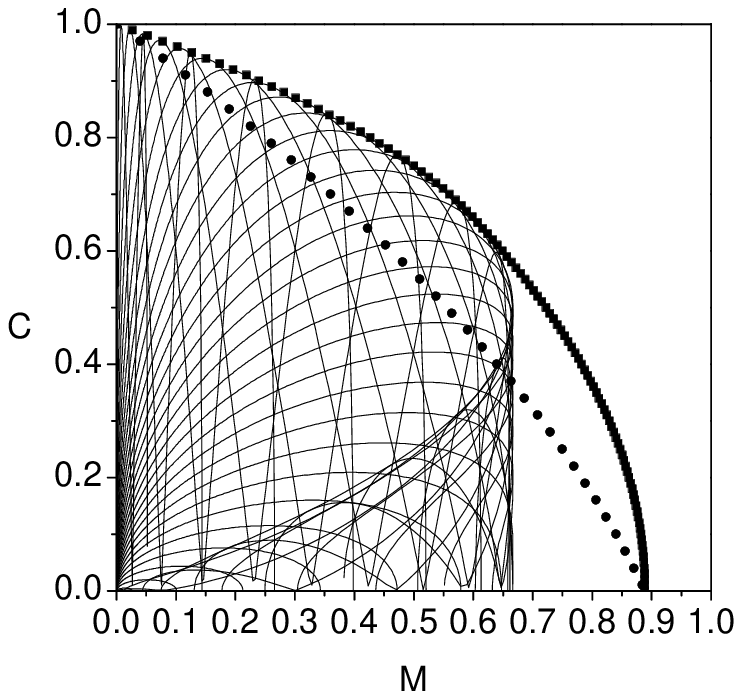}}
\caption{The concurrence versus mixedness of two atoms described
by the density matrix in Eq.(2) are depicted. Different
trajectories are chosen from the
$\frac{\lambda_2}{\lambda_1}\in[0,40]$ for different values of
${\Theta(t)}\in\{0.1,0.2,0.3,...,3.0\}$. The solid circle and
solid square represent the Werner state and the MEMS (the frontier
of the concurrence versus linear entropy) respectively. We can see
that part of the frontier of the concurrence versus linear entropy
can be reached by the evolving reduced density matrix of two
atoms.}
\end{figure}

Here, we confine our consideration in the MEMS in linear
entropy-concurrence plane. For the reduced density matrix $\rho_a$
in Eq.(2), the linear entropy is defined by
$M=\frac{4}{3}(1-{\mathrm{Tr}}\rho^2_a)$ which can be used to
quantify the mixedness of two atoms. In Fig.1, The concurrence
versus mixedness of two atoms are depicted for different values of
the coupling strength ratio. In each trajectory labelled by
different values of $\Theta{(t)}$, the data changes with
$\gamma_2/\gamma_1$. In this case, part of the frontier of the
concurrence versus linear entropy can be reached by the evolving
reduced density matrix of two atoms. It means that the MEMS can be
generated in the cavity QED system (1) if the ratio of coupling
strengths of two atoms is appropriately controlled. Not only pure
entangled states possessing any desired degree of entanglement can
be deterministically generated, but also the mixed states
possessing any possible degree of entanglement can be controlled
prepared if the desired linear entropy does not exceed a threshold
value near 0.65.

Next we discuss the Bell violation of two atoms in this system.
The most commonly discussed Bell inequality is the CHSH inequality
\cite{Bell,CHSH}. The CHSH operator reads \be
\hat{B}=\vec{a}\cdot\vec{\sigma}\otimes(\vec{b}+\vec{b^{\prime}})\cdot\vec{\sigma}
+\vec{a^{\prime}}\cdot\vec{\sigma}\otimes(\vec{b}-\vec{b^{\prime}})\cdot\vec{\sigma},
\ee where $\vec{a},\vec{a^{\prime}},\vec{b},\vec{b^{\prime}}$ are
unit vectors. In the above notation, the Bell inequality reads \be
|\langle\hat{B}\rangle|\leq2. \ee The maximal amount of Bell
violation of a state $\rho$ is given by \cite{Horodecki1995} \be
|{\mathcal{B}}|_{max}=2\sqrt{\kappa+\tilde{\kappa}}, \ee where
$\kappa$ and $\tilde{\kappa}$ are the two largest eigenvalues of
$T^{\dagger}_{\rho}T_{\rho}$. The matrix $T_{\rho}$ is determined
completely by the correlation functions being a $3\times3$ matrix
whose elements are
$(T_{\rho})_{nm}={\mathrm{Tr}}(\rho\sigma_{n}\otimes\sigma_{m})$.
Here, $\sigma_1\equiv\sigma_x$, $\sigma_2\equiv\sigma_y$, and
$\sigma_3\equiv\sigma_z$ denote the usual Pauli matrices. We call
the quantity $|\mathcal{B}|_{max}$ the maximal violation measure,
which indicates the Bell violation when $|{\mathcal{B}}|_{max}>2$
and the maximal violation when $|{\mathcal{B}}|_{max}=2\sqrt{2}$.
For the density operator $\rho_a$ in Eq.(2),
$\kappa+\tilde{\kappa}$ can be written as follows \be
\kappa+\tilde{\kappa}=C^2+\max[C^2,(1-\frac{\lambda^2_1(1-\cos2{\Theta(t)})}{\lambda^2})^2],
\ee

In Ref.\cite{Wei2003}, the analytical form of the mixed states
which possesses the maximal value of $|{\mathcal{B}}|_{max}$ of
two qubits for a given linear entropy has been derived. Now we
show that part of the frontier of the maximal Bell violation
versus the linear entropy can be approximately approached by two
atoms. In Fig.2, the Bell violation and linear entropy of the atom
1 and 2 are depicted. It can be observed that reduced density
matrix of two atoms can evolve into the vicinity of the frontier
of $|{\mathcal{B}}|_{max}$ versus $M$ if appropriate ratio of
coupling strengths and the time $t$ are chosen. Though the gap
between the MBVMS and the possible nearest reduced density matrix
becomes large when the linear entropy exceeds beyond 0.5, the
MBVMS with small mixedness can be approximately generated with
very high fidelity.

\begin{figure}
\centerline{\includegraphics[width=2.5in]{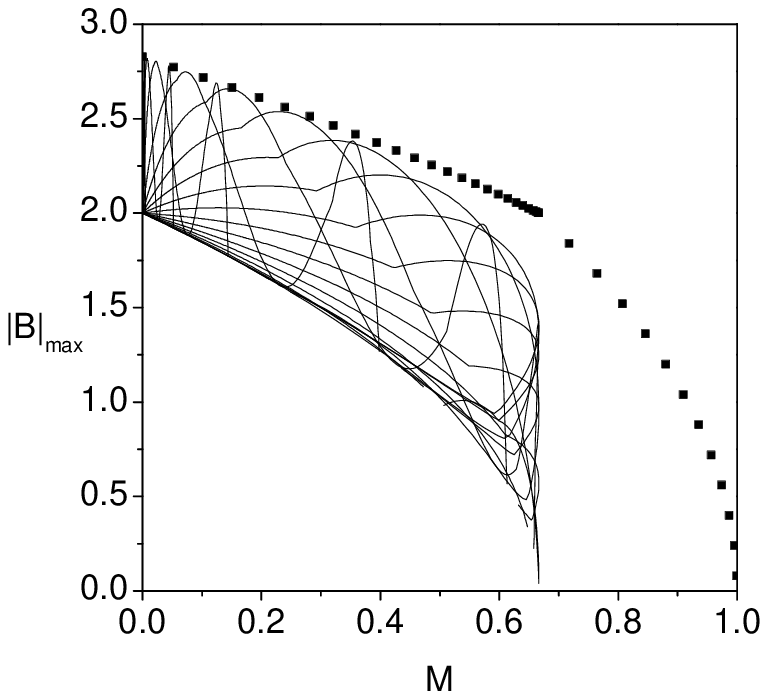}}
\caption{The maximal Bell violation $|B|_{max}$ versus the
mixedness of two atoms is displayed. Different trajectories are
chosen from the $\frac{\lambda_2}{\lambda_1}\in[0,40]$ for
different values of ${\Theta(t)}\in[0.1,3.0]$. The solid square
data points represent the frontier of maximal Bell violation
versus the linear entropy, namely, for a given linear entropy, the
maximal value of $|B|_{max}$ of two atoms can not exceed the solid
square data points.}
\end{figure}

\begin{figure}
\centerline{\includegraphics[width=3.5in]{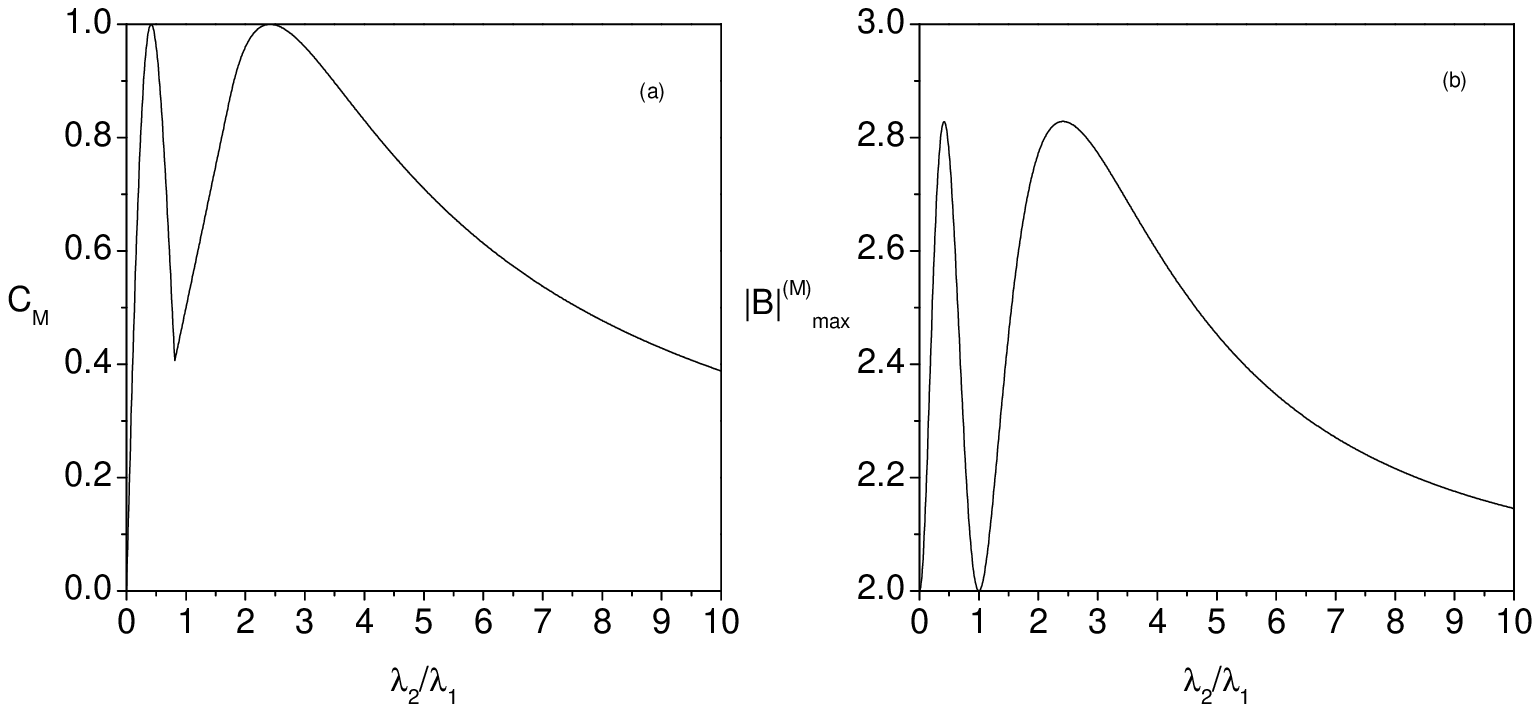}}
\caption{(a) The maximal value of the concurrence of two atoms
during the whole evolution is plotted as the function of the rate
of $\lambda_2/\lambda_1$. At $\lambda_2/\lambda_1=\sqrt{2}-1$ or
$\lambda_2/\lambda_1=\sqrt{2}+1$, two atoms can achieve the
maximally entangled state. It reflects the certain kind of partial
structural symmetry of the initial state. However, we can also
observer that the local minimum of the global maximal concurrence
is achieved at $\lambda_2/\lambda_1=\sqrt{\sqrt{32}-5}$ but not at
$\lambda_2/\lambda_1=1$, which is different with the case of
global maximum of $|B|_{max}$. (b) The maximal value of
$|B|_{max}$ of two atoms during the whole evolution is plotted as
the function of the rate of $\lambda_2/\lambda_1$.}
\end{figure}

In ref.\cite{Verstraete2002,Li20052}, it has been shown that
entanglement and Bell violation of two qubits may be inconsistent
with each other in certain kinds of dynamical processes. Here,
through investigating the global maximal values of concurrence and
Bell violation of two atoms initially in $|eg\rangle$, we obtain
more information about their dynamical discrepancy. From the
Eq.(5), one can easily derive the maximal value of the concurrence
of two atoms possibly achieved during the whole evolution as
follows: \be
C_M\equiv\max_{{t}\in[0,\infty)}C(t)=\frac{4|\chi-\chi^3|}{(1+\chi^2)^2},
\ee when ${\Theta(t)}=\pi$ if $0<\chi<\sqrt{\sqrt{32}-5}$ or
$\chi>\sqrt{3}$. Otherwise, \be C_M=\frac{\chi}{2}, \ee when
${\Theta(t)}=\arccos(\frac{1-\chi^2}{2})$ if
$\sqrt{\sqrt{32}-5}\leq\chi\leq\sqrt{3}$. In the above two
equation, $\chi\equiv\frac{\lambda_2}{\lambda_1}$. In deriving the
above or following maximal value, $\Theta(t)$ is assumed to be
large enough as $t\rightarrow\infty$. The maximum of the
concurrence of two atoms initially in $|eg\rangle$ completely
depends on the rate of $\chi$. Two atoms can become the maximally
entangled if $\chi=\sqrt{2}+1$ or $\chi=\sqrt{2}-1$
\cite{Olaya2004}. When two atoms symmetrically couples to the
cavity mode, the maximum of the concurrence is $\frac{1}{2}$. In
addition, in the cases with $1<\chi<\sqrt{3}$ or
$\frac{\sqrt{3}}{3}<\chi<1$, the maximum of the concurrence can
not keep invariant under the transformation of
$\chi\leftrightarrow\frac{1}{\chi}$, which implies that, in those
asymmetrically resonant coupled systems with
$1<\lambda_2/\lambda_1<\sqrt{3}$, two kinds of initial states
$|eg\rangle\otimes|0\rangle$ and $|ge\rangle\otimes|0\rangle$
result in the different maximum of the concurrence, namely, the
exchange symmetry is destroyed in this situation. However, for
other cases with $\chi\geq\sqrt{3}$ or
$\chi\leq\frac{\sqrt{3}}{3}$, the maximum of the concurrence can
keep invariant under the transformation of
$\chi\leftrightarrow\frac{1}{\chi}$ or the exchange of
$|eg\rangle\leftrightarrow|ge\rangle$. In Fig.3(a), we plot the
maximum of the concurrence as the function of the rate
$\lambda_2/\lambda_1$. It is shown that the maximum of the
concurrence $C_M$ of two atoms firstly increases from zero to $1$
with the increase of $\frac{\lambda_2}{\lambda_1}$ from zero to
$\sqrt{2}-1$, then decreases from 1 to
$\frac{\sqrt{\sqrt{32}-5}}{2}$ with the increase of
$\frac{\lambda_2}{\lambda_1}$ from $\sqrt{2}-1$ to
$\sqrt{\sqrt{32}-5}$. Furthermore, $C_M$ increases to $1$ when
$\frac{\lambda_2}{\lambda_1}$ increases from $\sqrt{\sqrt{32}-5}$
to $1+\sqrt{2}$, and then $C_M$ decreases with the further
increase of $\frac{\lambda_2}{\lambda_1}$. Further calculation
shows $\frac{\partial^2{C_M}}{\partial\chi^2}$ is discontinuous at
$\chi=\sqrt{3}$.

Then, we turn to investigate the global maximal value of Bell
violation $|{\mathcal{B}}|_{max}$ of two atoms during the whole
evolution. From the Eqs.(8,9), we can easily derive the analytical
expression for the maximal value of Bell violation of two atoms,
\be
|{\mathcal{B}}|^{(M)}_{max}\equiv\max_{{t}\in[0,\infty)}|{\mathcal{B}}|_{max}=2\sqrt{1+\frac{(4\chi-4\chi^3)^2}{(1+\chi^2)^4}},
\ee which is achieved by two atoms at the time ${\Theta(t)}=\pi$.
Surprisingly, the expression in Eq.(12) keeps invariant under the
transformation of $\chi\leftrightarrow\frac{1}{\chi}$ for the
whole range of the parameter $\chi$. In Fig.3(b), the global
maximum of Bell violation is plotted as the function of the rate
$\lambda_2/\lambda_1$. It is shown that
$|{\mathcal{B}}|^{(M)}_{max}$ exhibits the similar behavior to the
$C_M$ in most of the parameter $\chi$ except for
$\sqrt{\sqrt{32}-5}<\chi<1$, in which range $C_M$ increases but
$|{\mathcal{B}}|^{(M)}_{max}$ decreases.
\begin{figure}
\centerline{\includegraphics[width=3.5in]{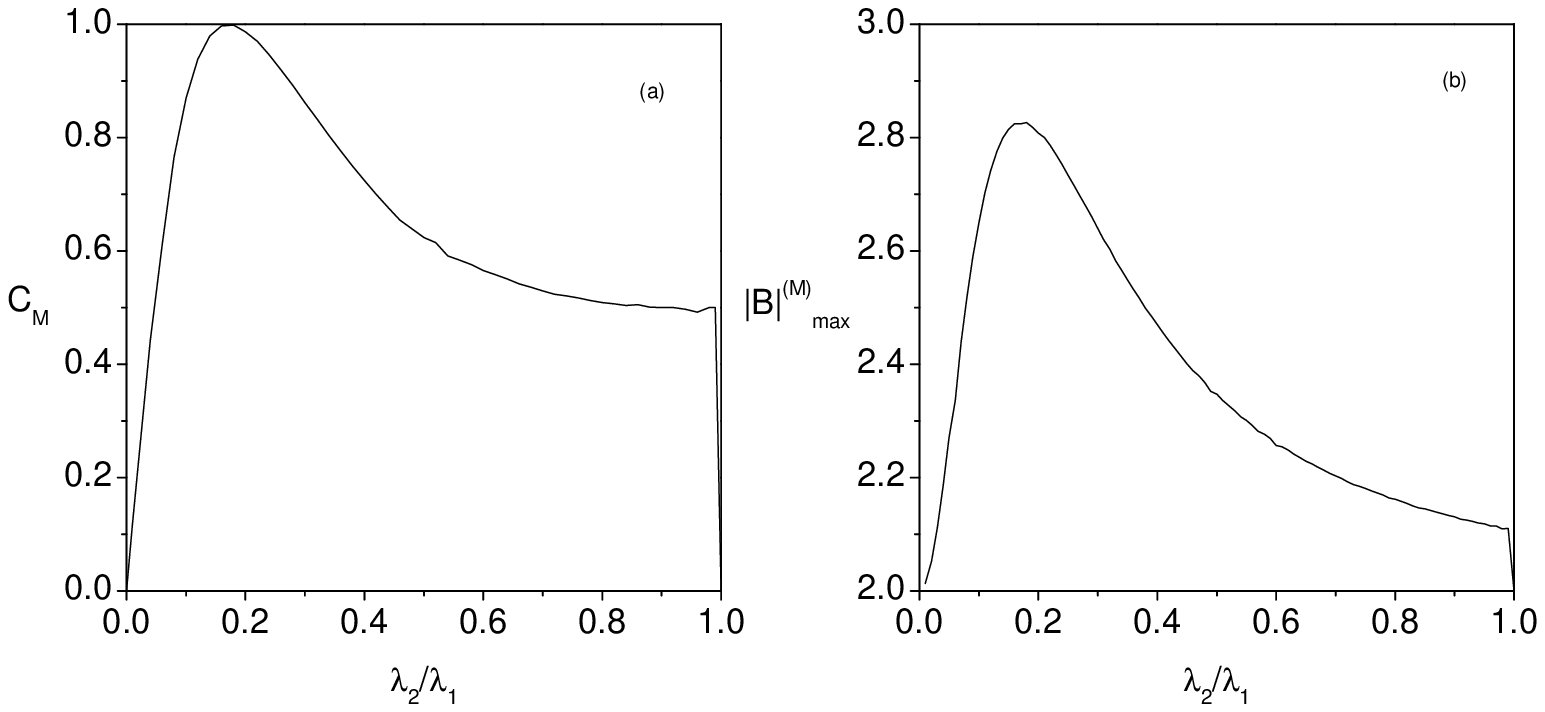}}
\caption{(a) The maximal value of the concurrence of two atoms
during the whole evolution is plotted as the function of the rate
of $\lambda_2/\lambda_1$. At $\lambda_2/\lambda_1\simeq0.18$, two
atoms initially in $|ee\rangle$ can achieve the maximally
entangled state, which also implies single-photon state of cavity
field can be simultaneously generated with the maximally entangled
state of two atoms. (b) The maximal value of $|B|_{max}$ of two
atoms initially in $|ee\rangle$ during the whole evolution is
plotted as the function of the rate of $\lambda_2/\lambda_1$.}
\end{figure}

The above analysis mainly focuses on the specific initial states,
i.e. the $|eg\rangle\otimes|0\rangle$ or
$|ge\rangle\otimes|0\rangle$. In what follows, we further consider
the case that two atoms are initially in the excited state
$|ee\rangle$ and the cavity field is still in the vacuum state
$|0\rangle$. In Fig.4(a), the maximal value of pairwise
concurrence of two atoms during the whole evolution is plotted as
the function of the coupling coefficients rate
$\lambda_2/\lambda_1$. It is shown that the maximal value of
pairwise concurrence firstly increases with $\lambda_2/\lambda_1$
and achieves 1 at $\lambda_2/\lambda_1\simeq0.18$, then decreases
with $\lambda_2/\lambda_1$. An abrupt decline from 0.5 to 0 can be
observed as $\lambda_2/\lambda_1\rightarrow1$. In Fig.4(b), the
global maximal value of $|{\mathcal{B}}|_{max}$ of two atoms
during the whole evolution is plotted as the function of the
coupling coefficients ratio $\lambda_2/\lambda_1$.
$|{\mathcal{B}}|^{(M)}_{max}$ exhibits similar dependence on
$\lambda_2/\lambda_1$ with the entanglement among all range of
$\lambda_2/\lambda_1$ in two-particle excitation case. A
interesting point is that we can simultaneously obtain pure
single-photon state in cavity field and a maximally entangled
state of two atoms in the situation of
$\lambda_2/\lambda_1\simeq0.18$. Both the single-photon state of
the cavity field and the maximally entangled state of two atoms
are the valuable resource in quantum information processes.
Therefore, it is hopeful that this scheme may have very
significant applications. One may conjecture that the influence of
spontaneous emission and cavity decay on the generation of
maximally entangled state in two-particle excitation case is more
significant than the one-particle excitation case. For fixing this
potential weakness, the timing single-photon detection may be a
good choice for purifying the entangled state of two atoms. The
details will be discussed elsewhere.

In summary, the generation of the maximally entangled mixed state
of two atoms which are asymmetrically coupled to a single mode
optical cavity field is analyzed. It is shown that two atoms can
achieve the maximally entangled mixed state or the maximal Bell
violating mixed states with high fidelity in the on-resonance
asymmetric coupling case. It is found that, for those cases with
$\frac{1}{\sqrt{3}}<\frac{\lambda_2}{\lambda_1}<\sqrt{3}$ and
$\lambda_1\neq\lambda_2$, exchange symmetry of global maximal
concurrence is broken when the initial state of two atoms
interchanges between $|eg\rangle$ and $|ge\rangle$. However, the
global maximum of maximal Bell violation keep invariant under the
exchange of two atoms. In the case of two-particle excitation, the
maximally entangled state of two atom can also be generated when
the coupling strength ratio is near 0.18. As the ratio of coupling
strengths tends to 1, the critical-phenomenon-like behaviors of
the global maximal entanglement or Bell violation can be found.
Though these results presented here can be directly generalized to
other systems such as two dipolar-coupled quantum dots in photonic
crystal microcavity, three-qubit Heisenberg spin chain, and
trapped ions coupled to motional degree of freedom etc, full
considerations of the effects of corresponding decoherence
mechanics on the preparation of MEMS or the partial
exchange-symmetry-breaking of global maximal entanglement are
nontrivial and very desirable.

The author thanks Prof. X.-B. Zou and Prof. J.-B. Xu for helpful
discussion.

\end{document}